\documentclass{article}

\usepackage[preprint]{neurips_2022}

\usepackage{natbib}

\usepackage[utf8]{inputenc} 
\usepackage[T1]{fontenc}    
\usepackage{hyperref}       
\usepackage{url}            
\usepackage{booktabs}       
\usepackage{amsfonts}       
\usepackage{nicefrac}       
\usepackage{microtype}      
\usepackage{xcolor}         
\usepackage{multirow}
\usepackage{amsmath}
\usepackage{graphicx}
\usepackage{floatrow}
\usepackage{blindtext}

\newfloatcommand{capbtabbox}{table}[][\FBwidth]

\DeclareMathOperator*{\argmin}{argmin}

\title{Physics-Driven Convolutional Autoencoder Approach for CFD Data Compressions}

\author{%
  Alberto Olmo \\ 
  National Renewable Energy Laboratory\\
  Golden, CO 80401 \\ 
  \texttt{aolmoher@asu.edu} \\
  \And
  Ahmed Zamzam \\
  National Renewable Energy Laboratory\\
  Golden, CO 80401 \\ 
  \texttt{ahmed.zamzam@nrel.gov} \\
  \And
  Andrew Glaws \\
  National Renewable Energy Laboratory\\
  Golden, CO 80401 \\ 
  \texttt{andrew.glaws@nrel.gov} \\
  \And
  Ryan King \\
  National Renewable Energy Laboratory\\
  Golden, CO 80401 \\ 
  \texttt{ryan.king@nrel.gov} \\
}

\begin{document}

\maketitle

\begin{abstract}
With the growing size and complexity of turbulent flow models, data compression approaches are of the utmost importance to analyze, visualize, or restart the simulations. Recently, in-situ autoencoder-based compression approaches have been proposed and shown to be effective at producing reduced representations of turbulent flow data. However, these approaches focus solely on training the model using point-wise sample reconstruction losses that do not take advantage of the physical properties of turbulent flows. In this paper, we show that training autoencoders with additional physics-informed regularizations, e.g., enforcing incompressibility and preserving enstrophy, improves the compression model in three ways: 
(i) the compressed data better conform to known physics for homogeneous isotropic turbulence without negatively impacting point-wise reconstruction quality,  
(ii) inspection of the gradients of the trained model uncovers changes to the learned compression mapping that can facilitate the use of explainability techniques, and
(iii) as a performance byproduct, training losses are shown to converge up to 12x faster than the baseline model.
\end{abstract}

\section{Introduction}
With the advancement of high performance computing (HPC) there has been an increase in the interest of leveraging such systems for computational fluid dynamics (CFD) simulations. These have become more readily available with larger than ever data sizes and performance fidelity~\citep{sprague2020exawind,fischer2021nekrs,musser2022mfix}. Further, recent advances in computational processing power achieved through heterogeneous architectures that couple traditional processors with graphics processing units (GPUs) have led to an increase in the gap between processing power and memory due to bandwidth constraints and memory-access times given by high latency input/output operations. This can lead to memory-bottleneck scenarios where HPC machines are limited by the need to save, analyze, visualize, or restore data from massive simulations. Given these factors and the increase in available datasets, it becomes critically important to develop in-situ data compression techniques to enable efficient use of the data without sacrificing accuracy. Additionally, a recent report from the U.S. Department of Energy (DoE), shows the analysis and visualization of CFD simulations as a central issue for next generation systems~\citep{DoE-vis}.

Several lossy compression approaches have been proposed recently \citep{Fukami2020AE-MSE, aeflow, AE-MSE-2, dunton2020pass} utilizing singular value decompositions or neural networks. In particular, convolutional autoencoders have proved to be able to obtain better generalization results~\citep{aeflow}. However, these approaches only leverage sample quality metrics while missing the physical properties inherent in the CFD and the benefits of embedding them at training time.
Therefore, motivated by the big successes of convolutional neural networks in the processing of CFD data~\citep{CNN-CFD-1, CNN-CFD-2} and in-situ data compression tasks~\citep{gan-compression}
and the increased ability to generate large CFD data sets, 
we develop a physics-driven convolutional autoencoder compression approach that builds on previous work~\citep{aeflow}. 

In this work, we show that using an autoencoder model enhanced with two physical properties of CFD leads to compression models that are more conformant with the physical characteristics of the data as measured by known metrics for homogeneous isomorphic turbulent flow, including the divergence-free condition of incompressible flow fields and the preservation of both enstrophy and dissipation ratio~\citep{constantin2020navier}. In addition, analyzing the model performance shows a significant reduction in training time as well as a reduced amount of training data necessary to achieve same-quality reconstructions as compared to the baseline. Further, our preliminary analysis of the learned models shows better explicability compared to models trained using only sample quality metrics.
All of this encourage the use of such networks with physical-losses and illustrate that gradient-based explainability techniques can be leveraged in the future. \footnote{The code of this work is attached as supplementary material and will be made publicly available.}


\section{Approach}

While lossless data compression approaches can be options for this problem~\citep{lossless1, lossless2}, they pose memory and execution time burdens for the system. On the other hand, lossy data compression aims to reduce the memory consumption by incurring some manageable loss of information after decompression. Hence, in our problem of reducing the dimensionality of CFD data and their memory expense, lossy compression methods are more suitable. Thus, a general lossy data compression function with full data space $\mathcal{X}$ and compressed data space $\mathcal{Y}$, can be defined in two parts: the compression step $\phi: \mathcal{X} \rightarrow \mathcal{Y}$ and reconstruction step $\psi: \mathcal{Y} \rightarrow \mathcal{X}$ where the degree of compression is measured by the compression ratio (CR).
To this end, we design a convolutional autoencoder with compression function $\phi$ defined by the encoder $E$, data $x$ and embedded data $z$ such that $E(x) = z$ and decompression function $\psi$ by the decoder $D$ such that $D(z) = \hat{x}$. Thus, the compressed data is obtained from $E(x)$ which can be stored more easily than the full data, and the full data reconstruction can be recovered with $D(z)$. 

\textbf{Data} The dataset we use consists of simulated snapshots of fluid velocities from incompressible decaying isotropic flows with component velocities on the $x$, $y$ and $z$ dimensions. Thus, the vector fields are comprised of 3-dimensional meshes of $128\times128\times128$ datapoints generated by the spectralDNS package~\citep{spectralDNS}. To increase robustness of the network, we introduce turbulences in the simulations as measured by Taylor-scale Reynolds numbers between (65, 105) and gather a total of 1300 snapshots for our training dataset.

\textbf{Physics-informed loss}
The network follows a fully convolutional architecture. In general, the main goal of a parameterized autoencoder is to minimize the reconstruction error with some pointwise metric such as the squared 2-norm:


\vspace{-8px}
\begin{equation}
\Theta_E, \Theta_D = \argmin_{\Theta_E, \Theta_D}\vert\vert x-D(E(x;\Theta_E) ; \Theta_D)\vert\vert^2_2 .
    \label{mse}
\end{equation}
\vspace{-8px}

Recent works have shown the advantages of using the physical properties of the domain during training. For instance, in \cite{pinns}, the authors train shallow neural networks with losses that include domain-specific physics laws and show improved generalization performance of the trained models. Similarly, \citet{pinns-review} show how physics-informed learning improves the inference performance for CFD domains such as three-dimensional wake flows or supersonic flows. Thus, in addition to MSE (\ref{mse}), we design the autoencoder loss with two physical laws that are applicable to the domain in consideration, the divergence-free condition and the preservation of enstrophy. For the former, due to the incompressibility of the flow field, the density of the CFD remains constant expressed by $\nabla \cdot \vec{v} = 0$ and therefore is a property that can be enforced. Similarly, the enstrophy of a fluid measures the kinetic energy in the flow that corresponds to dissipation effects and can be ensured to remain consistent between the original data and its reconstruction. Incorporating these properties into the training procedure, the loss becomes:
\begin{equation}
    \argmin_\theta \frac{1}{N}\sum_{i=1}^N (x_i-f(x_i;\theta))^2 + \frac{\lambda}{N}\sum_{i=1}^N(\nabla \cdot f(x_i;\theta))^2 + \frac{\beta}{N}\sum_{i=1}^N(g(x_i) - g(f(x_i;\theta))^2 ,
\end{equation}
where we denote a forward compression and decompression pass of the mesh $x_i$ on the network with parameters $\theta$ as $f(x_i; \theta) = D(E(x_i))$ and enstrophy $g(x)$ expressed in terms of the flow velocity as $g(x) \equiv \int_S \vert \nabla \times u \vert^2 dS$. We include the hyperparameters $\lambda$ and $\beta$ that allow tuning the sensitivity of the divergence-free minimization and preservation of enstropy respectively.

Secondly, we adapt the architecture from \cite{aeflow} in two ways. First, the model is adapted to handle simultaneous compressions of 3-channel velocity data in contrast to the one dimensional input of the former. Second, as wider layers are introduced to account for compressing this 3-velocity data, we adjust the network by removing its residual block layers. This proved to be an effective regularization method for the network and shows the network to be learning meaningful connections that the single-velocity baseline could have never inferred as we show next.


\begin{table}[t]
\centering
\scalebox{0.9}{
\begin{tabular}{cccccc}
\toprule
\multicolumn{2}{c}{Model} & \begin{tabular}[c]{@{}c@{}}Dissipation \\ Rate\end{tabular} & MSE & \begin{tabular}[c]{@{}c@{}}Mean \\ Divergence Loss\end{tabular} & \begin{tabular}[c]{@{}c@{}}Mean \\ Enstrophy Loss\end{tabular} \\ \midrule
\multicolumn{2}{c}{Vanilla ($\lambda=0$, $\beta=0$)} & 0.296 & 0.040 & 0.632 & 9.4e-5 \\ \midrule
\multirow{4}{*}{\begin{tabular}[c]{@{}c@{}}\\ \\ Divergence\\ $\lambda$ \end{tabular}} & 1e-2 & 0.476 & 0.047 & 0.080 & 10e-5 \\ \cmidrule{2-6} 
 & 1e-1 & 0.296 & 0.042 & 0.018 & 11e-5 \\ \cmidrule{2-6} 
 & \textbf{1} & 0.496 & 0.083 & \textbf{0.003} & 11e-5 \\ \cmidrule{2-6} 
 & 10 & 0.0001 & 0.967 & 3.6e3 & 6.9e5 \\ \midrule
\multirow{4}{*}{\begin{tabular}[c]{@{}c@{}}\\ \\ Enstrophy\\ $\beta$ \end{tabular}} & 1e-2 & 0.304 & 0.037 & 0.581 & 8.7e-5 \\ \cmidrule{2-6} 
 & 1e-1 & 0.411 & \textbf{0.036} & 0.664 & 9.6e-5 \\ \cmidrule{2-6} 
 & 1 & 0.304 & 0.037 & 0.578 & 8.2e-5 \\ \cmidrule{2-6} 
 & \textbf{10} & \textbf{0.504} & 0.066 & 0.682 & 7.5e-5 \\ \bottomrule \\
 \multicolumn{3}{c}{Ground truth dissipation rate: 0.544} & \multicolumn{3}{c}{Baseline model's MSE: 0.0604} 
\end{tabular}
}
\caption{\footnotesize Quantitative evaluations between our vanilla model and our model when trained with varying $\lambda$ and $\beta$. Each was trained with the same amount of data (1300 snapshots) and epochs (150). 
Ground truth dissipation rate is $0.544$ and closer values imply a better preservation. Baseline \citep{aeflow} MSE is $0.0604$ when trained with same data (only one channel at a time), epochs and learning rate.}
\label{tab:quantitative results}
\end{table}

\section{Experiments and Results}

We measure the improvements of our model by comparing it to the \textit{vanilla} version (when $\lambda=0$ and $\beta=0$) as well as the version from \citet{aeflow} (referred to as the \textit{baseline} version) in terms of quantitative and performance measurements which we outline below. 
For all experiments and models (both baseline and ours) we use the optimal baseline parameters as stated in \citet{aeflow}: 
a compression ratio of $CR=64$, batch size of 12, learning rate of $\eta = 1e^{-4}$ with scheduled decay and Adam optimizer~\citep{adam}. We train the models on 
a node from an HPC machine containing two NVIDIA Tesla V100 GPUs with 16 GB of dedicated memory and dual Intel Xeon Gold Skylake 6154 processors.

\textbf{Quantitative Experiments} We use four main evaluation metrics to assess the quality of the reconstructions over 1300 samples:  (i) the divergence of the flow field, (ii) the enstrophy mismatch loss, (iii) the pointwise mean squared error (as shown in \eqref{mse}), and (iv) the dissipation rate mismatch (given by $2\nu \langle S_ij S_ij\rangle$). We compile our results in Table \ref{tab:quantitative results}. Only one of $\lambda$ or $\beta$ was changed at each experiment while leaving the other hyperparameter at 0. From these results, we obtain several promising observations: our vanilla model ($\lambda=0$ and $\beta=0$) overperforms the baseline when trained under the same constraints in terms of the MSE reconstruction loss by a $33\%$ (0.04 vs 0.06). Further, the divergence loss of the new model is effectively reduced proportionally to higher $\lambda$ values (with the exception of $\lambda=10$ which makes the reconstruction loss explode) making the reconstructed snapshots more conformant with the divergence-free condition of the fluid data. On the other hand, the ground truth dissipation rate sits at $0.544$ and while training with $\lambda > 0$ values on average seem to yield its better preservation, our model achieves the closest at $\beta=10$ with $0.504$. It is worth noting that, overall, the inclusion of our physics terms during training shows to consistently improve the dissipation rate when compared to the vanilla as given by 6 out of the 8 models. 

Therefore, we show that the inclusion of the divergence-free and enstrophy-preserving terms to the loss prove not only to make the reconstructions more conformant with the physical laws of the domain but also these come at no significant expense in terms of reconstruction quality as measured by the MSE, even improving it in some cases.

\begin{figure}
\begin{floatrow}
\hspace*{-0.6cm}\ffigbox{%
  \includegraphics[width=0.92\columnwidth]{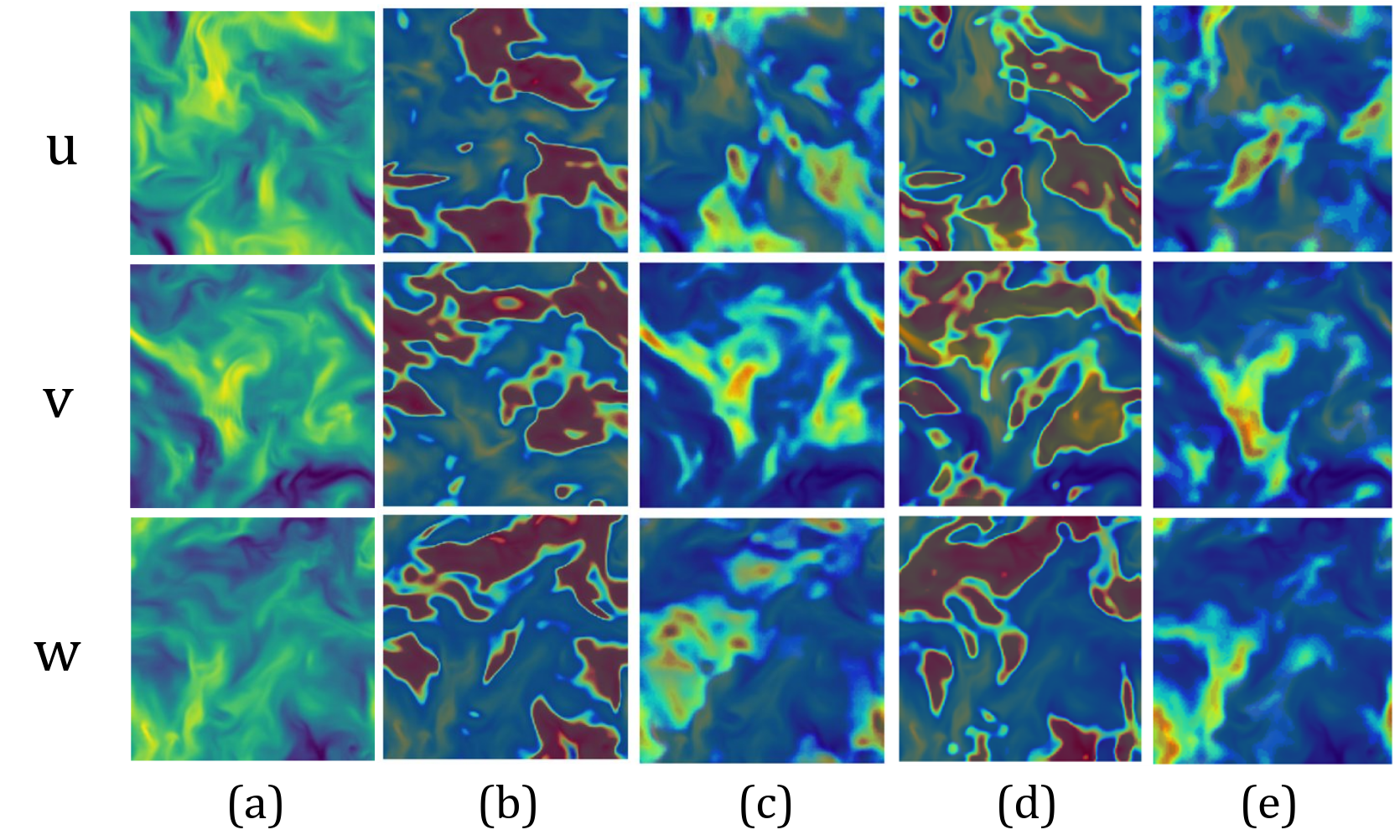}
}{%
  \quad\caption{\footnotesize Grad-CAM heatmaps over the $u$, $v$ and $w$ velocities on the learned gradients of the last layer of our model trained with $\lambda=0$, $\beta=0$ (c) $\lambda=1$ and $\beta=0$ (d) and $\lambda=0$ and $\beta=1$ (e). (a) and (b) columns represent the 3-channel snapshot data and the baseline's last layer gradients respectively.}%
  \label{fig:gradcam}
}
\capbtabbox{%
\scalebox{0.75}{
  \begin{tabular}{cccc}
\toprule
Model & \multicolumn{1}{l}{Training samples} & Training Time & MSE Loss \\ \midrule
Baseline & 1300 & 24h (50 epochs) & 0.045 \\
Ours & 1300 & \textbf{6h (50 epochs)} & \textbf{0.028} \\ \midrule
Baseline & 566 & 20.5h (100 epochs) & 0.045 \\
Ours & 566 & 5.5h (100 epochs) & \textbf{0.030} \\
Ours & 566 & \textbf{2.8h (50 epochs)} & 0.045 \\ \midrule
Baseline & 200 & 12h (180 epochs) & 0.052 \\
Ours & 200 & \textbf{54m (45 epochs)} & \textbf{0.045} \\ \midrule
\end{tabular}
}
}{%
  \caption{\footnotesize Training times and corresponding mean squared error loss for both the baseline model and ours with $\lambda=0$ and $\beta=0$ when trained with different amounts of data and epochs.}%
  \label{tab:performance results}
}
\end{floatrow}
\end{figure}

\textbf{Performance Experiments} We also assess the performance of the models in terms of training time by feeding and comparing both with same amounts of training samples (1300, 566 and 200 snapshots) and configurations. We report the results in Table~\ref{tab:performance results}. We note that our model significantly reduces the training time on average by a factor of 4 consistently over varying amounts of training data. At the same time, our model reaches better MSE loss when trained for the same number of epochs as shown in the first four sub-rows of Table~\ref{tab:performance results}. Thus, this allows the model to perform equally or better with fewer epochs or data as shown in the second row where ours gets the same reconstruction quality in half of the epochs (50) and a speedup of $7.3$x. It is worth noting that our model trains with 3-channeled data as opposed to the baseline which does one channel at a time, thus the speedup of the reported training times is further improved by a factor of 3 leading to up to a total of 12x speedup.

\textbf{Explainability Experiments} 
To examine the explainability of the reconstructions, we use a second order gradient-based technique that produces visual explanations via heatmaps from the gradient information of the last convolutional layer after a reconstruction~\citep{gradcam}. We show preliminary results in Fig.~\ref{fig:gradcam} and observe that the attention of the network changes depending on the regularization loss it was trained on. It is noticeable that the models trained using physics-informed losses focus on regions with higher magnitudes while the baseline models focus on random parts of the data. Due to this and the physical improvements of training with our loss, shown in Table \ref{tab:quantitative results}, we can infer that our showcased heatmaps become more focused on physically-relevant features and open up encouraging directions to further explore them due to the increasing need for explainability in deep learning for CFD.



\section{Conclusion and Future Work}

In this work, we tackle the problem of CFD data compression using autoencoders. Unlike recent literature focused on pointwise losses and motivated by the success of physics-informed constraints, we adapt and improve a baseline model to account for two physical laws of the CFD data that is trained on. We show that the benefits of the proposed approach are threefold: (i) the reconstructions effectively learn these laws, becoming more physics-conformant with the domain at no expense (or even improvement) in reconstructions quality, (ii) the adaptation of the model for 3-channel flow field data makes the network learn patterns between them allowing to reduce its depth, consequently speeding up its expensive training times significantly by a factor of up to 12x, and 
(iii) we apply gradient-based heatmaps on the last layer of the models and show that potentially more explicable patterns arise when trained with different weighted values of the physics-informed terms. This highlights pathways for future investigations to the interpretability of the compressions, and encourages the use of autoencoders and physics-informed losses as a potential candidate to tackle the rising interest in CFD models' explainability.

\bibliographystyle{plainnat}
\bibliography{ref}

\end{document}